\documentclass[12pt]{iopart}

\usepackage{graphicx}
\usepackage{hyperref}
\usepackage{amssymb}
\usepackage{color}
\usepackage{ulem}

\DeclareGraphicsExtensions{.eps}

\begin{document}

\title[Interpretation of experimental results on Kondo systems including crystal 
field effects]{Interpretation of experimental results on Kondo systems with crystal 
field}
\author{M. A. Romero$^1$, A. A. Aligia $^2$, J. G. Sereni$^2$ and G. Nieva$^2$}
\address{$^1$ Instituto de Desarrollo Tecnol\'{o}gico para la Industria
Qu\'{\i}mica (INTEC), U. N. del Litoral,  3000 Santa Fe, Argentina} 
\address{$^2$ Centro At\'omico Bariloche and Instituto Balseiro, Comisi\'on
Nacional de Energ\'{\i}a At\'omica, 8400 Bariloche, Argentina}

\date{\today }

\begin{abstract}
We present a simple approach to calculate the thermodynamic properties of single 
Kondo impurities including orbital degeneracy and crystal field effects (CFE) by  
extending a previous proposal by K. D. Schotte and U. Schotte 
[Physics Lett. A 55, 38 (1975)]. Comparison with exact 
solutions for the specific heat of a quartet ground state split into 
two doublets shows deviations below 10 \% in absence of CFE and a quantitative 
agreement for moderate or large CFE.
As an application, we fit the measured specific heat of the  
compounds 
CeCu$_2$Ge$_2$,  CePd$_{3}$Si$_{0.3}$, CePdAl, CePt, Yb$_2$Pd$_2$Sn and YbCo$_2$Zn$_{20}$. 
The agreement between 
theory and experiment is very good or excellent depending on the compound, 
except at very low temperatures due to the presence of 
magnetic correlations (not accounted in the model).

\end{abstract}


%75.20.Hr Local moment in compounds and alloys; Kondo effect, valence fluctuations, heavy fermions
%72.15.Qm Scattering mechanisms and Kondo effect
%75.40.Cx Static properties (order parameter, static susceptibility, heat capacities, critical exponents, etc.)

\maketitle

\section{Introduction}

\label{intro} In the last decades, heavy fermion systems and intermediate valence 
compounds have attracted considerable attention because of their fundamental importance 
in modern solid state physics  \cite{hews}. Different ingredients contribute to the 
complexity of these fascinating systems: the presence of strong Kondo interactions, 
the level structure originated in crystal-field effects (CFE) including 
different hybridization strengths with  the conduction band, and possible coherence 
effects introduced by the periodicity of the lattice and intersite magnetic interactions.

Usually above a certain coherence temperature, several Ce \cite{sereni,kroha} 
and Yb \cite{mon,zhou,moreno} compounds for example, can be described by an impurity Anderson model or the 
impurity Kondo model, which is the integer-valent limit of the former when
the magnetic configuration dominates. These systems behave as conventional 
Fermi liquids  \cite{Landau}, although with a very large effective mass ${{m}^{*}}$ 
\cite{Nozieres,Stewart}. Other systems
display non-Fermi-liquid behavior \cite{nfl}, but they are outside the scope
of this work.

The impurity Anderson model and several
variants of it, have been solved exactly using the Bethe ansatz technique 
\cite{Tsvelick,Andrei,bet,pedro,zpb,Desgranges,rasul,oki1,oki2,schlo}. 
Desgranges and Schotte have calculated
the specific heat for the spin 1/2 Kondo model solving numerically the
resulting system of integral equations \cite{Desgranges}. Exact results in presence of
crystal field have been reported by several authors \cite{rasul,oki1,oki2,schlo}. 
In particular, Desgranges and Schotte calculated
the specific heat of a system of two doublets \cite{rasul}.

A limitation of these exact solutions when CFE are present is that the 
hybridization of different multiplets are considered to be the same, 
which usually is not the case in real systems. Another drawback is that to calculate 
thermodynamic quantities, the Bethe ansatz leads to a set of integral equations, which 
should be solved numerically, rendering very difficult to use the exact solutions as a 
tool to fit experimental data. The numerical renormalization group (NRG) is a very accurate numerical 
technique which has the advantage over the Bethe ansatz that there
are no restriction for the impurity Hamiltonian \cite{bulla}. For example an Anderson model that 
mixes a doublet with either a singlet or o triplet can be exactly solved \cite{bet}, 
but not when both, singlet and triplet
are present, while the NRG can treat the complete model and describe the quantum phase transition
between singlet and doublet ground states \cite{allub}. However, the calculations 
are numerically demanding. To the best of our knowledge neither exact results nor NRG
has been used to fit experimental results in which non-trivial CFE are present.

Recent increase of the available experimental data of physical properties at 
intermediate temperature ($T \textgreater 10$K) on intermetallic compounds,
renders necessary the analysis of experimental data with the inclusion of CFE. 

In absence of crystal field, K. D. Schotte and U. Schotte 
proposed a simplified approach based on the resonant level model to interpret experiments 
of spin 1/2 Kondo systems in a magnetic field \cite{Schotte}. 
The theory assumes a linear increase of the slitting of the peaks with magnetic field,
which is only approximately true according to Bethe-ansatz results \cite{moore}. 
In spite of this fact, the authors were able to fit the 
magnetization of Fe impurities in Ag as a function of  magnetic field, and the magnetic susceptibility 
as a function of temperature very accurately \cite{Schotte}.  
For the specific heat, the comparison  between this approach with exact solutions 
shows an excellent agreement \cite{Desgranges}.

In this work we extend the approach to include CFE, and apply the results to interpret
the specific heat of several Ce and Yb compounds, for which one expects that single impurity behavior
in the Kondo limit (oxidation state near Ce$^{3+}$ or Yb$^{3+}$) and Fermi-liquid physics 
applies down to low temperatures. The approximations are described in Section 2.
The resulting approximate analytical results are compared with available exact results 
in Section 3. In Section 4 we apply our approach  
to four Ce (CeCu$_2$Ge$_2$,  CePd$_{3}$Si$_{0.3}$, CePdAl and CePt) and two Yb 
(Yb$_2$Pd$_2$Sn and YbCo$_2$Zn$_{20}$). 
compounds. 
We summarize our results in Section 5.

\section{Approximations}

\label{model}

We start from a generalization of the approach proposed by Schotte and Schotte \cite{Schotte} 
to the case of two doublets split by a crystal field $\Delta$ or, alternatively, a quadruplet 
with different $g$ factors, in both cases with the possible application of a magnetic 
field. Specifically we postulate the following simplified form of the free energy 
for two doublets 
\begin{equation}
{{F}_{2d}}=-{{k}_{B}}T\int\limits_{-\infty }^{\infty }{d\omega \frac{1}{\pi }%
\left[ \frac{{{\Gamma }_0}}{{{\left( \omega -\Delta_0\right) }^{2}}+\Gamma
_0^{2}}+\frac{{{\Gamma }_1}}{{{\left( \omega - \Delta_1 \right) }^{2}}%
+\Gamma _1^{2}}\right] \ln \left( {{e}^{\frac{\omega }{2{{k}_{B}}T}}}+{{e}%
^{-\frac{\omega }{2{{k}_{B}}T}}}\right) ,}  \label{f2d}
\end{equation}%
where ${\Gamma }_{i}$ represents the half width at half maximum of the spectral density
for the  doublet $i$, $\Delta_1=\Delta +s$ where $s \le \Gamma_0$ 
is a small shift that can be disregarded for the moment, 
$\Delta_0=B_0+s$ and $B_0=g_0\mu _{B}\tilde{B}$  is the Zeeman magnetic splitting 
of the doublet ground state by the magnetic field $\tilde{B}$.
The Kondo temperature $T_K$ is proportional to the width of the ground-state doublet,
$\Gamma_0$, as explained in detail in the next section. 
Note that in absence of hybridization effects  (i.e. ${{\Gamma}_{i}}=0$), except for an irrelevant additive 
constant, ${{F}_{2d}}$ reduces to

\begin{equation}
F_{2d}^0=-k_B T \ln \left( e^{\frac{B_0/2}{k_B T}}+ e^ {-\frac{B_0/2}{k_B T}}
+e^{\frac{\Delta +B_0/2}{k_B T}}+e^{\frac{\Delta -B_0/2}{k_B T}}\right) ,  \label{f2d0}
\end{equation}
which corresponds to the free energy of two doublets separated by an energy $\Delta$, 
both split by a magnetic field in the same fashion. Although the latter fact is not realistic,  
for usual cases in which  $\Delta \gg B_0$, the  individual levels of the excited doublet 
have a similar population for all temperatures and their magnetic splitting becomes irrelevant.
The same situation occurs for  ${{\Gamma }_1}\gg B_0$. 
Thus we expect that Eq. (\ref{f2d}) is a reasonable approximation in general.

An alternative scenario  to which Eq. (\ref{f2d}) can be applied is a quadruplet split by a 
magnetic field. In that case the corresponding  energies are $\pm
g_{0}\mu _{B}\tilde{B}/2$ and $\pm g_{1}\mu _{B}\tilde{B}/2$ and  one should consider 
${{\Gamma }_{1}={\Gamma }_{0}}$, $\Delta_1 =(g_{0}+g_{1})\mu
_{B}\tilde{B}/2$, $\Delta_0=(g_{0}-g_{1})\mu _{B}\tilde{B}/2$.

The integral in Eq. (\ref{f2d}) with a cutoff $D$ can be expressed in terms
of the Gamma function $\Gamma \left( x\right)$:
\begin{eqnarray}
{{F}_{2d}} &=&2{{k}_{B}}T\mathbf{Re} \sum_{j=0}^1\left\{ \ln \Gamma \left( 1+\frac{{{\Gamma }_j}
+i\Delta_j }{\pi {{k}_{B}}T}\right) -\ln \Gamma \left( 1+\frac{{{\Gamma }_j}
+i\Delta_j }{2\pi {{k}_{B}}T}\right) \right\}  \nonumber \\
&-&\left( \frac{{{\Gamma }_{0}}+{{\Gamma }_{1}}}{\pi }+2{{k}_{B}}T\right)
\ln 2+\frac{{{\Gamma }_{0}}+{{\Gamma }_{1}}}{\pi }\left( 1-\ln \frac{D}{{{k}
_{B}}T}\right)  \label{f2db}
\end{eqnarray}

In practice, we find that for small bare splitting $\Delta$, it is convenient to increase
the position of both doublets by a shift $s$ with $0 < s \le \Gamma_0$. 
This can be understood as follows.
For the SU(2) model, the approximation of Schotte and Schotte seems to be inspired by the fact
that the electron spectral density has a resonance centered at the Fermi energy, in such a way that in 
the Kondo limit, with total electron occupation 1 at the impurity, both spins have occupation 1/2.
This becomes clear in Fermi liquid approaches, in which the resonance near the Fermi energy
can be accurately described \cite{hew,scali}. However, when the hybridization between the impurity
and conduction electrons is the same for both doublets and $\Delta=0$, the model has SU(4) 
symmetry \cite{lim,ander,desint,buss,su42,see,tetta}, each fermion should have an occupation near 1/4, 
and this means that the resonance is displaced above the Fermi energy \cite{su42,see}.
For general hybridizations, and zero magnetic field, integrating the spectral densities 
(in a similar way as in the slave-boson mean-field approximation \cite{see}), one finds that
to enforce a total occupation 1 
(as corresponding to Ce and Yb impurities in the Kondo limit) 
in a mean-field approach, the shift $s$ should satisfy 

\begin{equation}
\frac{2}{\pi}[{\rm arctan}\frac{\Gamma_0}{s}+{\rm arctan}\frac{\Gamma_1}{s+\Delta}]=1,
\end{equation}
which is equivalent to
\begin{equation}
\Delta_0 \Delta_1=\Gamma_0 \Gamma_1,
\label{shift}
\end{equation}
whre we have used that $\Delta_0=s$ for $B=0$. The extension of this heuristic approach to finite magnetic 
field is beyond the scope of the present work.
In any case the above equation shows that $s$ is negligible for large $\Delta$.

In Ce compounds, the Hund's rules ground state with total angular momentum $j=5/2$ 
is split in a quartet and a doublet, 
or three doublets depending on the point group determined by the symmetry around the Ce atom. 
Therefore, for a comparison with experiment, a third doublet should be included. 
Note that including a third term between brackets in Eq. (\ref{f2d}) 
would mean that eight (not six) broadened levels are considered (as it is clear taking all ${{\Gamma }_{i}}=0$). 
Therefore, a straightforward extension of this equation to include three doublets is not possible. 
However, if the width of the levels ${{\Gamma }_{i}}$ is much smaller than the splitting $\Delta_2$ 
between the third and the first doublet, we can take at high temperatures
the simple Schottky expression $F_{S}$, which corresponds to all ${{\Gamma }_{i}}=0$:

\begin{eqnarray}
F_{S} &=&-{{k}_{B}}T\ln \left( 2c_{0}+2c_{1}{{e}^{-\frac{\Delta_1 }{{{k}_{B}}T}}}
+2c_{2}{{e}^{-\frac{\Delta_2}{{{k}_{B}}T}}}\right) ,  \nonumber \\
c_{i} &=&\cosh (g_{i}\mu _{B}\tilde{B}/2).  \label{fs}
\end{eqnarray}%
A similar reasoning can be followed for more doublets.
This leads to the following proposal to describe the free energy for three
doublets, or a quartet ground state and an excited doublet. 
\begin{equation}
F=F_{2d}-F_{2d}^{0}+F_{S}  \label{f}
\end{equation}
For small temperatures ${{F}_{2d}}$ dominates, while the remaining two terms become important 
when ${{k}_{B}}T$ reaches values near $\Delta_2$. This will become apparent in Section \ref{meas}. 
Note that this form ensures the correct limit $\left( {{k}_{B}}\ln 6\right) $ for the entropy as 
$T\rightarrow \infty $. For a ground state doublet and an excited quadruplet, ${{F}_{2d}}$ and ${{F}_{2d}^{0}}$ 
should be replaced 
by the corresponding expressions for one doublet in which the terms containing $\Delta_1$ are absent. 
Extension of Eq. (\ref{f}) to include more doublets (like e.g. in the case of Yb systems with $J=7/2$)  
with width much smaller than its excitation energy is straightforward.

The specific heat and magnetic susceptibility of the system can be obtained differentiating 
Eq. (\ref{f}) \cite{Schotte}. Here for later use, we give the expression of the specific heat at 
zero magnetic field ($\tilde{B}=0$)

\begin{equation}
C={C}_{2d}-\frac{1}{{k}_{B}{T}^{2}}\left
[\frac{{{\Delta }^{2}}{{e}^{-\frac{\Delta }{{{k}_{B}}T}}}}
{{{\left( 1+{{e}^{-\frac{\Delta }{{{k}_{B}}T}}}\right) }^{2}}}\right] + C_S ,
\label{c}
\end{equation}
where the specific heat for only two doublets or a quartet 
${C}_{2d}=-T\partial ^{2}F_{2d}/\partial T^{2}$ becomes

\begin{eqnarray}
 C_{2d} &=& -\frac{k_B}{2 \left( \pi k_B T \right)^2}
\mathbf{Re} \sum_{j=0}^1 \left\{ 
\left( \Gamma_j+i\Delta_j \right)^2 \left[ 4\psi ^{\prime }\left( 
\frac{\Gamma_j+i\Delta_j }{\pi k_B T}\right) -\psi ^{\prime
}\left( \frac {\Gamma _2+i\Delta }{2\pi k_B T} \right) \right] \right\} 
\nonumber \\
&& + \frac{\Gamma_0 + \Gamma_1}{\pi T},  \label{c2d}
\end{eqnarray}%
where $\psi ^{\prime }$ is the derivative of the digamma function,
and $C_S$ is the Schottky expression for the specific heat obtained deriving Eq. (\ref{fs}):

\begin{equation}
C_S=\frac{1}{{k}_{B}{T}^{2}}\left[ \frac{{{\Delta_1 }^{2}}{{e}^{-\frac{
\Delta_1 }{{{k}_{B}}T}}}+{{\left( \Delta_2 \right) }^{2}}{{e}^{
-\frac{\Delta_2 }{{{k}_{B}}T}}}+{{\left( \Delta_2 -\Delta_1 \right) }
^{2}}{{e}^{-\frac{\Delta_2+\Delta_1 }{{{k}_{B}}T}}}}{{{\left( 1
+{{e}^{-\frac{\Delta_1 }{{{k}_{B}}T}}}+{{e}^{-\frac{\Delta_2}{{{k}_{B}}T}}}
\right) }^{2}}} \right].
\label{cs}
\end{equation}

For later use, we give here the coefficient $\gamma$ of the first term in the low-temperature
expansion $C= \gamma T + O(T^2)$ of the specific heat:

\begin{equation}
\gamma=\frac{\pi k_B^2}{3}\sum_{j=0}^1 \frac{\Gamma_j}{\Delta_j^2+\Gamma_j^2}.
\label{gam}
\end{equation}

\section{Comparison with exact solutions}

\label{compa}

While several works discuss the solution of the Bethe-ansatz equations with CFE \cite{rasul,oki1,oki2,schlo},
to the best of our knowledge, the solution in a wide range of temperatures has been reported
only by Degranges and Rasul \cite{rasul}. These authors calculated the specific heat as a 
function of temperature for a quartet split into two doublets by an energy $\Delta$ and 
zero magnetic field. For $\Delta=0$, this Kondo model has SU(4) symmetry, which is reduced
to SU(2) as soon as $\Delta>0$. This SU(4) $\rightarrow$ SU(2) model, 
and the corresponding (more general) Anderson one, has become popular 
recently in the context of nanoscopic systems \cite{lim,ander,desint,buss,su42,tetta,see}. 
It describes for example quantum dots in 
carbon nanotubes,\cite{lim,ander,buss,su42,see} and silicon nanowires \cite{tetta,see} in presence of a
magnetic field and interference effects in two-level systems \cite{desint,benzene}

Before comparing the results of Ref. \cite{rasul} with ours, it is convenient to discuss
briefly the meaning of the Kondo temperature in systems with CFE.

\subsection{The Kondo temperature}

\label{kt}

We define the Kondo temperature $T_K$ as the binding energy of the ground-state singlet,
as done for example in a perturbative-renormalization-group study of the Kondo model with 
CFE \cite{yama} or variational (non-perturbative) methods \cite{moreno,su42}. Alternative 
definitions, like for example from the width of the lowest peak in the spectral density 
\cite{su42,see},  temperature dependence of transport properties \cite{see,cornut}, or the linear term 
in the specific heat \cite{rasul} give 
the same result within a factor of the order of 1, but the leading exponential dependence
on the parameters is the same. In the extreme limits in which either no CFE are present,
or they are so large that only the ground-state multiplet matters, the Kondo temperature
is given by the usual expression for an SU(N) Kondo model

\begin{equation}
T_K^{SU(N)}=D {\rm exp} (-2 \rho J/N),  
\label{tksun}
\end{equation}
where N is the degeneracy of the ground state, $D$ half the band width, $J$ the exchange interaction, 
and $\rho$ the conduction density of states at the Fermi level.

For the case of a quartet split into two doublets (the SU(4) $\rightarrow$ SU(2) model mentioned above),
using a simple variational function, it has been found that the Kondo temperature as a function
of the splitting $\Delta$ can be written as \cite{su42}

\begin{eqnarray}
\frac{T_K(\Delta)}{T_K(0)}=\sqrt{1+\delta/d +\delta^2}-\delta, \nonumber \\
\delta=\frac{\Delta}{2 T_K(0)} {\rm ,~   } d=\frac{D}{2 T_K(0)}
\label{tk},
\end{eqnarray}
where of course, $T_K(0)=T_K^{SU(4)}$, and from the above equations $T_K(\infty)=T_K^{SU(2)}$.

Using the non-crossing approximation, the width of the peak nearest to the Fermi energy of
the spectral density has been found to be accurately described by Eq. (\ref{tk}) within a 
constant factor near 0.6 \cite{su42}.

\subsection{The specific heat for two doublets}

\label{dr}

%\begin{figure}[tbp]
\begin{figure}[!ht]
\begin{center}
\includegraphics[width=9.0cm]{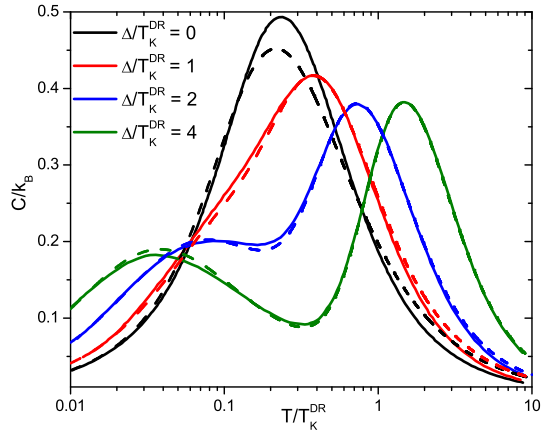} \\[0pt]
\end{center}
\caption{(Color online) Impurity contribution to the specific heat for a Kondo model with two doublets. 
Full lines correspond to exact results of Ref. \cite{rasul}, dashed lines to our approach. The absolute
maximum displaces to higher temperatures with increasing $\Delta$.}
\label{fig1}
\end{figure}

The exact Bethe-ansatz solution of the specific heat of the Kondo model for a quartet split into two 
doublets has been reported by Degranges and Rasul \cite{rasul}.
In Fig. (\ref{fig1}) we compare their results with our simplified proposal for the specific heat, 
Eq. (\ref{c2d}). The authors define their Kondo temperature $T_K^{DR}$ by the condition that 
in the SU(4) case $\Delta=0$, the term linear in temperature $T$ in the specific heat for
$T \rightarrow 0$ is $\gamma=k_B \pi/T_K^{DR}$. In our case SU(4) symmetry and Eq. (\ref{shift})
imply $\Delta_0=\Delta_1=\Gamma_0=\Gamma_1$. For these parameters, Eq. (\ref{gam}) gives 
$\gamma=k_B \pi/(3 \Gamma_1)$.
Comparing this result with the corresponding one of Degranges and Rasul fixes $\Gamma_0=T_K^{DR}/3$.
Except for this, there are no fitting parameters for zero splitting in our approach. Taking into account 
this fact, we find that the agreement is very satisfactory. Our approach underestimates the maximum in the 
specific heat by about 10\%, but the overall trend and the low-temperature part are well reproduced. 
The comparison is however much better when the splitting $\Delta$ reaches $T_K$ or higher values. 
In fact, for large  $\Delta$, the lower peak corresponds to the ordinary Kondo model with one doublet, 
for which the approach works very well \cite{Desgranges}.

As soon as $\Delta>0$, we use three parameters to fit the exact results: both $\Gamma_i$ and $\Delta_1$, with 
the position of the lower doublet $\Delta_0=s$ determined by the condition (\ref{shift}). As discussed below, 
the resulting parameters are consistent with expectations from known results on the the SU(4) $\rightarrow$ SU(2)
symmetry breaking. The fitting parameters are displayed in Table \ref{para}, together with the corresponding 
shift.

\begin{table}[t]
\caption{\label{para} Parameters $\Gamma_i$, $\Delta_1$ used in the fit of Fig. \ref{fig1} in units of $T_K^{DR}/3$.
The first column corresponds to the Kondo temperature given by Eq. (\ref{tk}) and the last to
the shift determined by Eq. (\ref{shift}). }
\begin{tabular}{llllll}
$\Delta$ & $0.6 T_K$ & $\Gamma_0$ & $\Gamma_1$ & $\Delta_1$ & $\Delta_0$ \medskip \\ \hline
%\vspace*{-10pt}
\\
0 &  1/3 &  1/3 &  1/3 &  1/3 &  1/3  \medskip \\
1 & 0.173  & 0.237  & 0.4 & 0.94 & 0.1009   \medskip \\
2 & 0.118  &  0.14  & 0.47 & 1.75 & 0.0376   \medskip \\
4 & 0.0806  & 0.078  & 0.55 & 3.49 &  0.0123  \medskip \\
\end{tabular}
\end{table}

The first column  of Table \ref{para} 
is an estimate of $\Gamma_0$ based on the expected dependence of the Kondo temperature 
with splitting, Eq. (\ref{tk}). Results using the non-crossing approximation show that the width of the peak
near the Fermi energy in the spectral density is proportional to $T_K$ for all $\Delta$ \cite{su42}. 
The proportionality 
factor for the charge transfer energy used was found to be 0.606. 
We assume $\Gamma_0=0.6 T_K$. Using this relation
for $\Delta=0$, the ratio $T_K^{SU(4)}/T_K^{DR}$ is obtained, and for the other values of $\Delta$,
Eq. (\ref{tk}) was used, with a band width $2D=10 T_K^{DR}$. Larger values of $D$ lead to lower values of $T_K$ 
but affect little the values for $\Delta \ge 4 T_K^{DR}$. As seen in Table \ref{para}, the estimate of
the width of the lowest peak in the spectral density agrees with $\Gamma_0$ within 30\%, while the values of 
$\Gamma_0$ for different values of $\Delta$ vary by more than a factor 4. This is a strong indication
that the value of $\Gamma_0$ that results from the fit is proportional to the binding energy 
of the ground-state singlet.

Concerning $\Gamma_1$, one observes a moderate increase as $\Delta$ increases, a fact also shared by
the spectral density studied before \cite{su42}. One way to understand this fact is to consider the spectral 
density of the simplest SU(2) Anderson model in the Kondo regime, under an applied magnetic field $B$ (which
for the peak at higher energies, is a simpler analog of the SU(4) model under a magnetic or crystal field). 
Clearly, the width of the 
resonance is proportional to $T_K^{SU(2)}$ for $B=0$. As $B$ increases, the Kondo effect is 
progressively destroyed and for very large $B$ one expects that the width of the peak in the spectral density 
is just the resonant level width, which is larger than $T_K$. This is consistent with 
Bethe-ansatz results \cite{moore}. These results also show that the position of the peak is lower
than the magnitude of the Zeeman term, but tends to it for large $B$, in agreement with the fact that
our fit gives $\Delta_1 < \Delta$ for $\Delta>0$. Calculations of the spectral density in 
the SU(4) $\rightarrow$ SU(2) case, shows peaks above and below the Fermi energy 
(depending on the component), with excitation energies smaller than $\Delta$ \cite{su42}, a fact also
consistent with the fitting results.

\section{Application to real systems}

\label{meas}

In this section, we apply our approach to interpret the specific heat measured on four Ce
and two Yb compounds at zero applied magnetic field. 
These systems have an oxidation state of the magnetic atom near Ce$^{3+}$ or Yb$^{3+}$ 
(they are in the Kondo regime) and seem to display Fermi-liquid single-ion behavior down to low temperatures.
The Ce compounds (CeCu$_2$Ge$_2$,  CePd$_{3}$Si$_{0.3}$, CePdAl and CePt with increasing Kondo temperature $T_K$) 
order at low temperatures. Above this ordering temperature, one expects that the system
can be described by a generalized impurity Kondo model, for which our approach was developed.
The Yb compounds that we considered (Yb$_2$Pd$_2$Sn and YbCo$_2$Zn$_{20}$) do not order down 
to very low temperatures.

As it is known, the possibility of ordering depends on the magnitude of the exchange interaction $J$
between localized spins and conduction electrons.
For small $J$ the indirect Ruderman-Kittel-Kasuya-Yosida (RKKY) exchange interaction 
between localized spins is much larger than $T_K$ and the system orders, while
for large $J$, single-ion Kondo physics dominates down to zero temperature and the system does not order.
This is the basis of the so called Doniach's phase diagram \cite{don}. For only one doublet,
the critical ordering temperature $T_N$ as a function of $J$ has been calculated using the exact
magnetic susceptibility of the impurity system and treating the RKKY interaction at a 
mean-field level \cite{lobo}. $T_N$ first increases quadratically with $J$, reaches a maximum and then 
decreases until it vanishes at a quantum critical point $J=J_c$. Near this point obviously
$T_N \ll T_K$, whereas for very small $J$, $T_N \gg T_K$. In general we expect that our theory
is valid for $T > T_N$, even if $T_N > T_K$ as for the case of CeCu$_2$Ge$_2$ discussed below.

\subsection{CeCu$_2$Ge$_2$}
\label{Ge}

The structure of CeCu$_2$Ge$_2$ is tetragonal. Compounds with this structure were intensively
investigated after the discovery of superconductivity in CeCu$_2$Si$_2$ \cite{ste}.
The specific heat of the Ge compound is reported in Fig. 1 of Ref. \cite{felten}. 

\begin{figure}[!ht]
\begin{center}
\includegraphics[width=9.0cm]{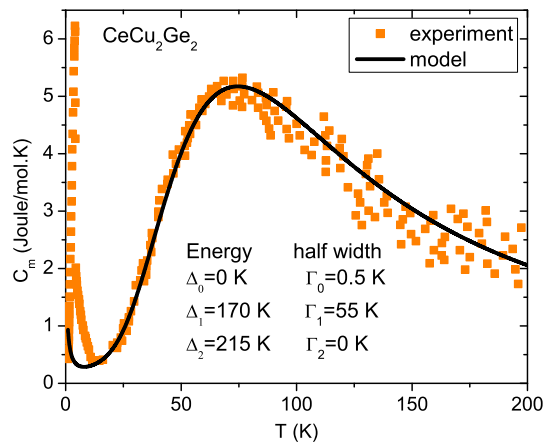} \\[0pt]
\end{center}
\caption{(Color online) Magnetic contribution to
the specific heat as a function of temperature of CeCu$_2$Ge$_2$. 
Squares: experimental results \cite{felten}. Full line: our approach.}
\label{fge}
\end{figure}

In Fig. \ref{fge} we fit the magnetic contribution to specific heat data. 
The tetragonal symmetry, in particular the 
point group symmetry $D_{4h}$ around the Ce$^{3+}$ ions, implies that the  $j=5/2$ ground state multiplet 
splits into three doublets.
Thus, we can apply Eqs. (\ref{c}), (\ref{c2d}) and (\ref{cs}). In principle, our expression for the specific heat 
has four fitting parameters 
($\Gamma_{0}$, $\Gamma _{1}$, $\Delta_1$ and $\Delta_2$). The shift $\Delta_0=s$ calculated with 
Eq. (\ref{shift}) is very small and does not 
provide any significant change in the fit. We have neglected it for the all the Ce compounds studied.
In practice however, we have we used only the $\Gamma_i$ as fitting parameters, taking 
the same values of $\Delta_i$ as in the 
Schottky expression with three doublets, Eq. (\ref{cs}),
provided in Ref. \cite{felten} to compare with
the experimental results. This expression overestimates the peak near 70K by about 20\%
(see Fig. 1 of Ref. \cite{felten}) and underestimates the specific heat between 12 and 25 K.
Instead, as it is apparent in Fig. (\ref{fge}) above 12 K (where the experimental curve has a kink), 
our fit is excellent and the difference 
between experiment and the figure is less than the experimental error. Thus, the introduction
of the widths $\Gamma _{0}$ and $\Gamma _{1}$ by our approach leads to a significant improvement
of the fitting expression with almost the same computational cost.

The system orders antiferromagnetically at $T_N=4.15$ K. Below 12 K, our fit deviates from experiment.
This is probably due to the onset of antiferromagnetic correlations between  Ce$^{3+}$ ions which
can not be captured in an approach based on a single impurity like ours. We have a similar limitation
in the fits described below. 
Since the data at very low temperatures deviates from experiment, the reader might ask, 
how sensitive the fit is
to changes in $\Gamma _{0}$ which, as discussed in section \ref{compa}, is proportional to
the Kondo temperature $T_K$. If $\Gamma _{0}$ is increased from 0.5 K to 1 K, the specific heat $C$
increases by about 15\% in the region between 12 and 25 K. If $\Gamma _{0}=2$ K, $C$ nearly doubles for
$T=12$ K and lies above the experimental data for $T <30$ K.

\subsection{CePd$_{3}$Si$_{0.3}$}
\label{Si}

\begin{figure}[!ht]
\begin{center}
\includegraphics[width=9.0cm]{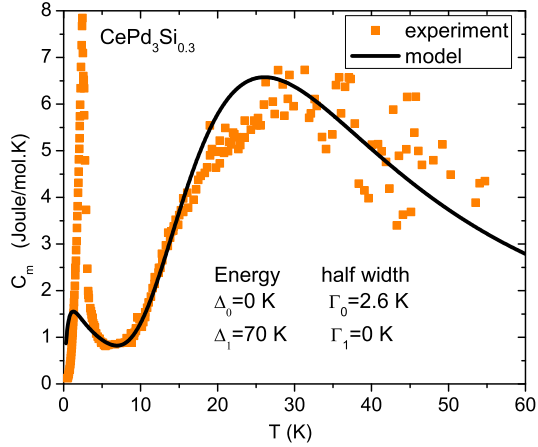} \\[0pt]
\end{center}
\caption{(Color online) Magnetic contribution to
the specific heat as a function of temperature for CePd$_{3}$Si$_{0.3}$. 
Squares: experimental results \cite{nieva2}. Full line: our approach.}
\label{fsi}
\end{figure}

CePd$_{3}$ has a cubic structure and behaves as an intermediate valence system. 
Doping with B or Si expands the lattice and drives the system
to the Kondo regime \cite{nieva}. To fit the specific heat, we have chosen CePd$_{3}$Si$_{0.3}$,
because 
%the effects of disorder (which are beyond the scope of this work) are expected
%to be less than with B replacement, and 
the subtraction of the phonon contribution has been done by 
measuring the specific heat of a La compound with a similar Si content, LaPd$_{3}$Si$_{0.2}$.
The specific heat of both compounds has been measured by one of us \cite{nieva2}. 
The low temperature behavior for LaPd$_{3}$Si$_{0.2}$ has been fit with an $aT+bT^3$ 
dependence to obtain the Debye temperature, of a simple Debye model for the phonon
contribution to the specific heat \cite{ash}. Then, the Debye expression for 
this contribution is subtracted from the 
corresponding data for CePd$_{3}$Si$_{0.3}$, to obtain the magnetic contribution $C_m$ to
the specific heat of this compound. The result is represented in Fig. (\ref{fsi}) together with
our fit. 

Since the point group around a Ce ion is $O_h$, we have used a ground state doublet and an excited quadruplet 
for the fit, as described in Section 2. Since we are not able to introduce a width to the excited quadruplet
within our approach, we have only two fitting parameters, the width of the ground state doublet
$\Gamma _{0}$ and the position of the quadruplet $\Delta_1$. The fit is very good for temperatures between
5 K and 18 K. For $T < 5$ K, as for the other compounds, the disagreement with experiment is due to the 
effects of magnetic correlations. Above 30 K, the dispersion of the experimental data is too large
to allow us to extract firm conclusions, but the agreement is reasonable. Instead in the intermediate regime
18 K$<T<$ 27 K the fit lies above the experimental data. This is likely due to the lack of broadening of
the quadruplet in our approach. Some effects of disorder due to random distribution 
of Si atoms at the center of the cubic cage built by eight Ce$^{3+}$ ions, might also play a role.  

\subsection{CePdAl}
\label{Al}

\begin{figure}[!ht]
\begin{center}
\includegraphics[width=9.0cm]{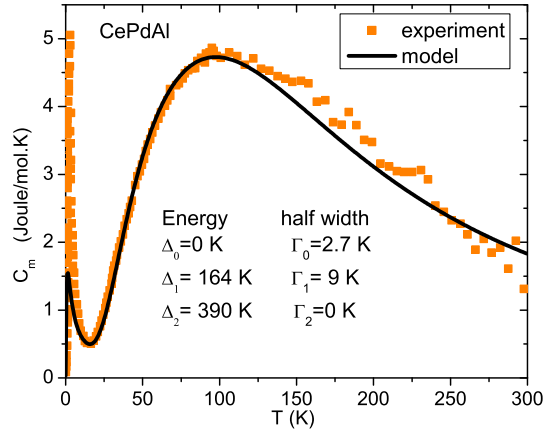} \\[0pt]
\end{center}
\caption{(Color online) Magnetic contribution to
the specific heat as a function of temperature for CePdAl. 
Squares: experimental results \cite{woit}. Full line: our approach.}
\label{fal}
\end{figure}

CePdAl is a heavy-fermion system which crystallizes in the hexagonal ZrNiAl structure and 
orders antiferromagnetically at $T_N=2.7$ K \cite{woit}. The specific heat is reported in Fig. 4
of Ref. \cite{woit} together with a fit using a Schottky expression with three doublets, Eq. (\ref{cs}). 
This fit falls
clearly below the experimental data for $T<50$ K. 

Our fit using  Eqs. (\ref{c}), (\ref{c2d}) and (\ref{cs}) is displayed in
Fig. (\ref{fal}) together with the experimental data. At $T > 130$ K, where the experimental data
seem to have more dispersion, our fit falls slightly below the experimental data, which nevertheless show 
more dispersion at high temperatures. The fit fails of course
near $T_N$ because it is based on a single ion approach, which cannot describe magnetic order.
At intermediate temperatures $T_N < T <130$ K, our fit is excellent. 

Considering the whole range of temperatures, our fit is again superior to the Schottky expression
with a similar computational effort.

\subsection{CePt}
\label{Pt}

The crystal symmetry of CePt is orthorhombic. Thus, it is another material in which the Ce$^{3+}$ ions
lie in low symmetry sites and the $j=5/2$ multiplet is split into three doublets. The magnetic contribution to 
the specific heat $C$ has been measured in Ref. \cite{bur}.  The data together with a Schottky calculation 
[Eq. (\ref{cs})] are 
presented in Fig. 2 of this reference. As for CeCu$_2$Ge$_2$, this expression overestimates $C$ near the 
peak at about 75 K, and underestimates it below 30 K. There is also a funny structure between 250 K and 300 K 
in which the experimental data lie above the Schottky expression. 

\begin{figure}[!ht]
\begin{center}
\includegraphics[width=9.0cm]{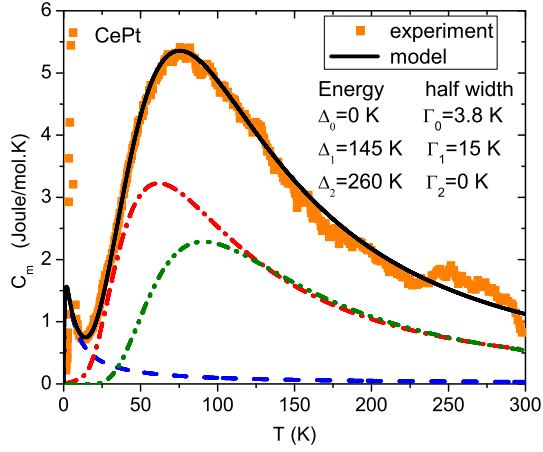} \\[0pt]
\end{center}
\caption{(Color online) Magnetic contribution to
the specific heat as a function of temperature for CePt. 
Squares: experimental results \cite{bur}. Full line: our approach.
Dashed, dot-dashed and dot-dot-dashed curves copprespond to the contribution of
the different doublets.\cite{note} (see text).}
\label{fpt}
\end{figure}

The data are represented in
Fig. \ref{fpt} together with our fit using Eqs. (\ref{c}), (\ref{c2d}) and (\ref{cs}). As before, the values 
of $\Delta_1$ and 
$\Delta_2$ were chosen the same as those proposed in Ref. \cite{bur} from the fit using the
Schottky expression and only two $\Gamma_i$ were varied.
The fit is excellent except near or below the ordering temperature
(6.2 K) and near the above mentioned high-temperature structure. 

We also display the contributions proportional to $\Gamma_{0}$, $\Gamma_{1}$ and the remaining term, 
which can be interpreted loosely \cite{note} as the contribution of each doublet to $C$.
Specifically, the dashed (dot-dashed) line corresponds to the terms containing
$\Gamma_0$ ($\Gamma_1$) in the expression of $C_{2d}$ [Eq. (\ref{c2d})], 
while the dot-dot-dashed curve is the correction 
due to the third doublet and corresponds to the
two remaining terms of Eq. (\ref{c}). 

\subsection{Yb$_2$Pd$_2$Sn}
\label{YbPd}

No magnetic transition is found above $T=0.5$ K in this compound \cite{kik} and therefore it is believed
to be a non-magnetic compound with $J > J_c$ in the Doniach's phase diagram \cite{lobo}.
It crystallizes in a tetragonal structure. Therefore, the $j=7/2$ ground state of Yb is split into 
four doublets. Then, we can describe its specific heat with Eqs. (\ref{c}) and (\ref{c2d}), with Eq. (\ref{cs})
generalized to contain an additional doublet at energy $\Delta_3$. 

In contrast to the above Ce compounds, in both Yb compounds that we considered, the splitting of
the first excited doublet is not small compared to the widths of the two lowest lying levels, and therefore the 
shift $s$ (which coincides with $\Delta_0$ in absence of magnetic field) cannot be neglected. In fact,
using $\Delta_0$ given by Eq. (\ref{shift}) leads to a substantial improvement of the fits 
as compared to $\Delta_0=0$.

The magnetic contribution to the specific heat (after subtracting the specific heat of 
Lu$_2$Pd$_2$Sn) has been reported by Kikuchi {\it et al.} \cite{kik}. The authors also show in their Fig. 2 
an interpretation of the data based on the sum of a Kondo 1/2 contribution plus a crystal-electric-field 
contribution with three doublets, which clearly fall much below the data near 30 K. They auggest that taking
the Kondo effect of excited states into account would improve the description of the data. This is indeed what 
we have done for the first excited state.

\begin{figure}[!ht]
\begin{center}
\includegraphics[width=9.0cm]{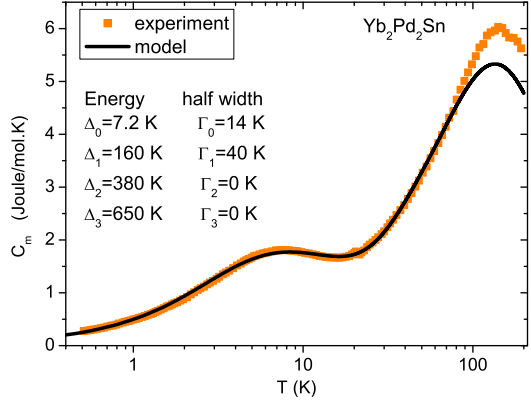} \\[0pt]
\end{center}
\caption{(Color online) Magnetic contribution to
the specific heat as a function of temperature for Yb$_2$Pd$_2$Sn. 
Squares: experimental results \cite{kik}. Full line: our approach.}
\label{fybpt}
\end{figure}

The comparison between these experiments and our theory is shown in Fig. \ref{fybpt}. 
The values of the different $\Delta_i$ are consistent with inelastic neutron scattering data of
a similar compound Yb$_2$Pd$_2$In with In instead of Sn \cite{bauer}.
The agreement is very good except above 100 K, where some entropy is lacking in our approach.
Presumably this is due to the broadening of the two higher levels, particularly the highest one, which 
are absent in our approach. These broadenings would contribute to the entropy at smaller temperatures
than the position of the respective peaks at $\Delta_2$ and $\Delta_3$.
In spite of this shortcoming, our fit is much better to the curve presented in Fig. 2 of Ref. \cite{kik}
in the whole temperature range.

\subsection{YbCo$_2$Zn$_{20}$}
\label{YbZn}

\begin{figure}[!ht]
\begin{center}
\includegraphics[width=9.0cm]{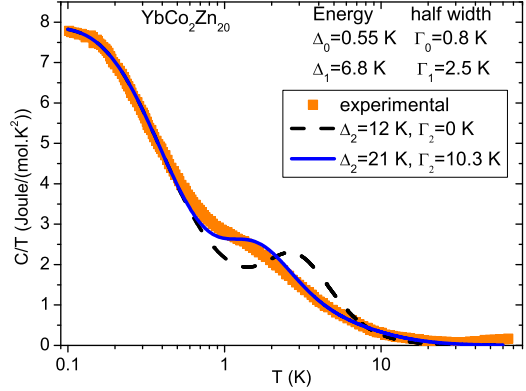} \\[0pt]
\end{center}
\caption{(Color online) Magnetic contribution to
the specific heat divided by temperature as a function of temperature for YbCo$_2$Zn$_{20}$. 
Squares: experimental results \cite{take}. Dashed line: our approach. Full line:
modification to simulate a width of the quadruplet (see text).}
\label{fybco}
\end{figure}

This heavy fermion compound is characterized by an extremely large value of $C/T$ at low temperatures.
The magnetic contribution to the specific heat has been measured by Takeuchi {\it et al.} \cite{take} 
and the results are reproduced by square symbols
in Fig. \ref{fybco}. The compound is cubic, and therefore the $J=7/2$ states of Yb are split into two doublets
($\Gamma_6$ and $\Gamma_7$) and a $\Gamma_8$ quadruplet. Our fit indicates that the latter lies at higher energy.
Following our approach we used then Eqs. (\ref{c}) and (\ref{c2d}), but with $C_S$ replaced by the corresponding
term obtained differentiating the free energy $F_S$ for all $\Gamma_i=0$, which in this case is
\begin{eqnarray}
F_{S} =-{{k}_{B}}T\ln \left( 2+2{{e}^{-\frac{\Delta_1 }{{{k}_{B}}T}}}
+4{{e}^{-\frac{\Delta_2}{{{k}_{B}}T}}}\right).   
\label{fs4}
\end{eqnarray}
The resulting fit is shown by the dashed line of Fig. \ref{fybco}. 
Clearly, the fit is very good at temperatures below 1 K, but fails between 1 and 10 K, slightly 
above the energy of the quadruplet. We believe that the cause of this discrepancy is the lack of broadening
of this quadruplet in our theory. To test this, we have replaced the last term of Eq. (\ref{fs4})
by a sum of four terms with equidistant energies $E_k=\Delta_2 -\Gamma_2(k-3/2)$, $k=0$ to 3, which simulates
the effect of the broadening (although naturally the queues of the distribution are lost).
This improves the experimental curve, except at small temperatures as described above.

In contrast, the agreement at low temperatures, in which the experimental $C/T$ is rather flat, 
is a success of our approach. The shift $\Delta_0$ given by Eq. (\ref{shift}) is essential to reproduce 
this behavior. Using $\Delta_0=0$ leads to a too fast decay of $C/T$ with increasing temperature.

Concerning the extremely high value of the linear term of the specific heat $\gamma=7.8$ 
Joule/(mole K$^2$), our result using Eq. (\ref{gam}) and the parameters of the fit (given in the figure)
indicate that 5.3 \% of the observed $\gamma$ is due to the first excited doublet.

\section{Summary and discussion}

\label{sum}

We have presented a simple approach to describe the thermodynamics of
Kondo impurities with several levels, extending a previous proposal
of Schotte and Schotte \cite{Schotte}. 
In absence of an applied magnetic field, a comparison with available exact results
indicates that the approach describes properly the specific heat of systems with
two doublets, and accurately as long as the splitting is equal or larger than the Kondo temperature.
The parameters of the fit are related to fundamental quantities
which describe the spectral density of the system. In particular $\Gamma_0$ is related 
to the Kondo temperature.

Application to specific heat measurements
of several  compounds with low symmetry around the Ce$^{3+}$ ions  
(CeCu$_2$Ge$_2$,  CePdAl and CePt)
provides an excellent fit at temperatures above those at which 
magnetic correlations between different Ce$^{3+}$ ions becomes important.
For the cubic system CePd$_{3}$Si$_{0.3}$, our fit deviates above the
experimental results in a narrow range of intermediate temperatures.
This might be due to the limitations of our approach, which does not include
a broadening of excited quadruplets, or to effects of disorder.
In the case of the Yb compound YbCo$_2$Zn$_{20}$, in which the position of the excited 
quadruplet is of the order of its broadening, the effect of the lack of the latter
becomes evident in the fit. An artificial splitting of this quadruplet improves the fit, 
but this does not represent the actual physics of the hybridization of
excited states with conduction electrons.
In contrast, the low-temperature part, which shows an unusually flat $C/T$ is very well reproduced
by our approach.
Finally for Yb$_2$Pd$_2$Sn, which has the largest $T_K$ and a crystal-field splitting
of four doublets, our fit is very good except at high energies, where it is likely 
that the effect of broadening of excited states also plays a role.

For all the above systems our fit was superior to alternative theoretical curves
if available, and the computational cost is very modest.

We conclude that the present approach provides a flexible tool to properly interpret 
experimental results, at temperatures larger than those corresponding to the onset of 
either coherence effects of the lattice 
or magnetic correlations.

\section*{Acknowledgments}

We thank CONICET from Argentina for financial support. This work was
partially supported by PIP 112-200801-01821 and PIP 112-00621 of CONICET, 
and PICTs R1776 and 2007-812 of the ANPCyT, Argentina.

{\it Note added.} Recently we became aware of the work in Ref. \cite{de2}, in which Degranges 
extends the excat Bethe ansatz results of Ref. \cite{rasul} to the case of three
doublets, all of them with the same hybridization to the conduction states.


\begin{thebibliography}{99}
\bibitem{note} The partition function and not the free energy is additive in terms of the 
contributions of different states which mutually exclude each other. Therefore, the contribution 
of each doublet to the specific heat is not well defined.
\end{thebibliography}
\end{document}